\documentclass{iau}
\usepackage{graphicx,natbib}
\usepackage{hyperref}

\title{Rings of star formation:\\Imprints of a close galaxy encounter}

\author[J. Moreno]{{Jorge Moreno$^{1,\dagger,2}$}}

\affiliation{
$^{1}$Department of Physics and Astronomy, University of Victoria, Finnerty Road, Victoria, British Columbia, V8P 1A1, Canada\\
$^{2}$Department of Physics and Astronomy, California State Polytechnic University, Pomona, CA 91768, USA\\
$^{\dagger}$CITA National Fellow\\
e-mail: {\tt jorgemoreno@csupomona.edu} 
}

\pubyear{2014}
\volume{309}
\jname{Galaxies in 3D across the Universe}
\editors{B. L. Ziegler, F. Combes, H. Dannerbauer, M. Verdugo, eds.}

\begin{document}

\maketitle

\begin{abstract}
In this talk, I report results from galaxy merger simulations, which suggest the existence of a ring of star formation produced by close galaxy encounters. This is a generic feature of all galaxy interactions, provided that the disc spins are sufficiently aligned. This signature can be used to identify close galaxy pairs that have actually suffered a close interaction.
\keywords{galaxies: galaxies: formation -- evolution -- interactions}
\end{abstract}

\begin{figure}
\centering
\includegraphics[width=\columnwidth]{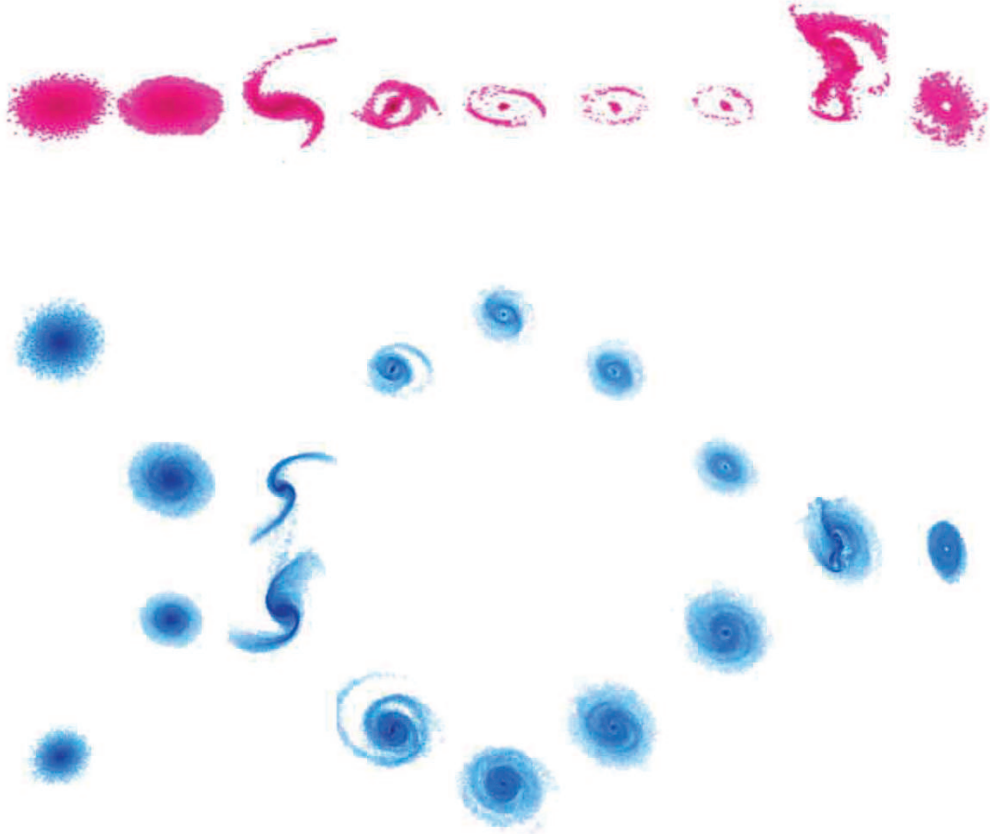} 
\caption{Interacting galaxies with strongly-aligned disc-spin orientation. {\it Left-to-right:} Various stages of interaction and merging: incoming, close pericentric passage, receding, re-approaching, and merging. {\it Top:} The morphology of star formation in the secondary (smaller) galaxy (in pink). {\it Bottom:} The gas-morphology of the two interacting galaxies (in blue -- zoomed-out scale).}\label{fig:fig1}
\end{figure}

\firstsection
\section{Introduction}

Over thirty five years ago, \cite{larson78} first recognised interactions as a promising avenue for triggering star formation in galaxies. This has been a subject of intense work by numerical simulators \citep{barnes91,barnes96,cox06,dimatteo07,dimatteo08,cox08}.  With the help of large surveys, such as the Sloan Digital Sky Survey, interactions are now established as a prime driver for igniting star-forming episodes \citep{ellison08,ellison10,ellison11,patton11,scudder12,patton13}, active galactic nuclei \citep{ellison11,ellison13b}, and for the creation of tidal tails in galaxies with close companions \citep{casteels13}.

Indeed, it is common practice to use the presence of tidal tails as an indicator of past interaction \citep{kartaltepe12,hung13,hung14}. However, a problem with this approach is that for many orbits, tidal tails may dissipate long before the two galaxies merge together. Moreover, some orbital configurations may be more conductive to creating tidal tails than others. In other words, it could be the case that visual classification (via the presence of tails) might miss a fraction of truly interacting galaxy pairs. 

In this presentation, I report a signature produced by galaxy interactions: a ring of star formation. This feature may serve as an alternative to identifying galaxies that have experienced a close encounter.  A more exhaustive study of this ring-like structure is reserved for future work (Moreno et al., in prep). 

\section{Methods, Results \& Discussion}
I employ a suite of 75 SPH (smoothed-particle hydrodynamics) merger simulations \citep{springel05gadget}, comprised of three disc-spin orientations selected from \cite{robertson06}: the ``e", ``f", and ``k" orientations -- meant to represent aligned, perpendicular, and anti-aligned spins, respectively. See \cite{torrey12} and Moreno et al. (in prep) for details. For each of the three orientations considered, we focus on five eccentricities ($\epsilon=\{0.85, 0.90, 0.95, 1.0, 1.05 \}$) and five impact parameters ($b=\{2, 4, 8, 12, 16\}$ kpc).

Figure~\ref{fig:fig1} shows the results for one of our orbits ($\epsilon=1.05$, $b=16$ kpc, ``e" orientation). In blue (bottom), we show the morphology of the gas component for both galaxies. From left-to-right, each stage of the merger sequence is presented: incoming (1-2), first pericentric passage (3), receding (4-5), re-approaching (6-7), merging (8-9). Tidal tails are only evident briefly (3). In pink (top), we show the morphology of star formation (zoomed-in scale). After first pericentric passage, tidal tails wrap around into a ring of off-nuclear dense gas. This accumulation of material triggers a stable star-forming ring-like structure, which survives for a prolonged period.

This feature takes $\sim$0.2-1 Gyr after first pericentric passage to appear. It is triggered for all orbits in the ``e" and ``f" orientations (aligned and perpendicular spins), regardless of $\epsilon$ and $b$. However, when the disc spins are anti-aligned (``k" orientation), this ring-like feature is absent (with the exception of a few anomalous orbits where the two disc interpenetrate one another).


\begin{figure}
\centering
\includegraphics[width=\columnwidth]{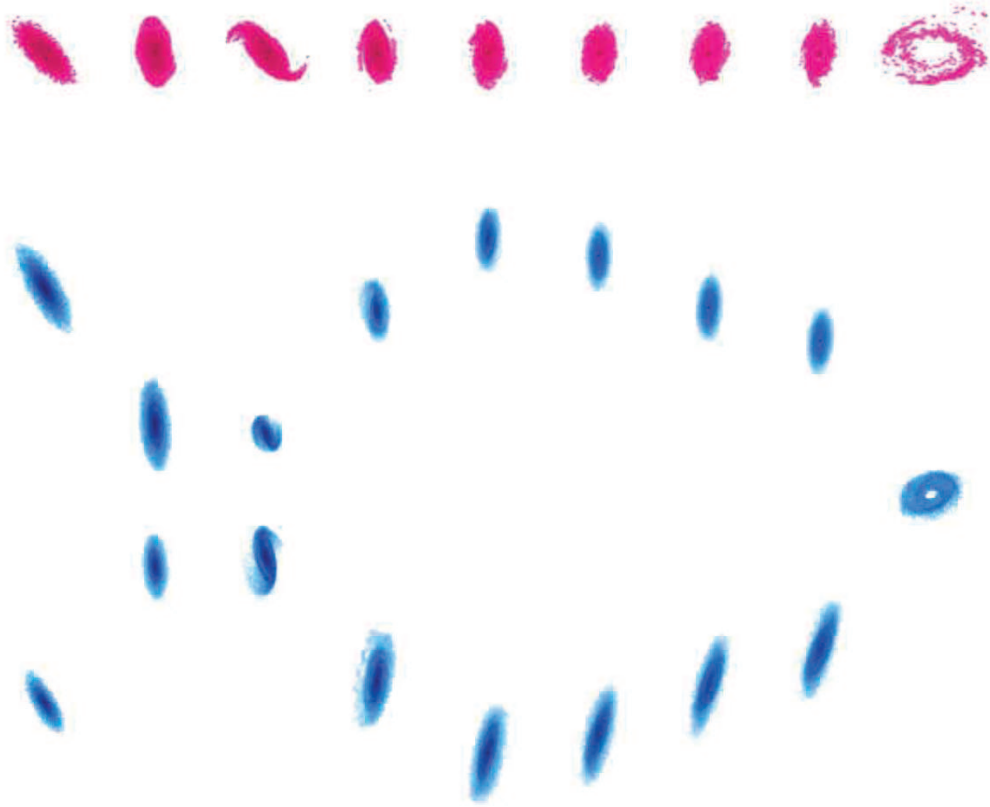} 
\caption{Interacting galaxies with anti-aligned orientation. Same format as in Figure~\ref{fig:fig1}.}\label{fig:fig2}
\end{figure}

\section{Conclusions}

In this talk, I report the existence of ring-like off-nuclear star formation produced by close galaxy interactions. If the two discs are sufficiently aligned, a ring-like feature is triggered on the secondary galaxy soon after first passage.

In principle, this is a promising way of identifying those galaxy pairs that have actually experienced a close encounter in the past -- which, in turn, would allow more refined calculations of the galaxy merger rate \citep{patton97,bluck12,lopezsanjuan13} -- but see \cite{moreno12} and \cite{moreno13} for caveats.

One could argue that, if observed, there might be other causes for the existence of this ring. For instance, a ring-like structure is detected in M31 \citep{gordon06}.  However, based on its off-centre nature, simulations suggest that this was caused instead by a direct off-nuclear collision \citep{block06,dierickx14}.

It is my hope that these findings motivate observational investigations by integral-field spectroscopic surveys, such as \textsc{califa} \citep{sanchez14}, \textsc{sami} \citep{croom12}, \textsc{m}{\small a}\textsc{nga} (Bundy et al., in prep), and the future \textsc{hector} survey \citep{lawrence12}.

\section*{Acknowledgements}

\noindent
The computations in this paper were run on the Odyssey cluster supported by the FAS Division of Science, Research Computing Group at Harvard University. I acknowledge the Natural Science and Engineering Research and the Canadian Institute for Theoretical Astrophysics for funding. This presentation would not have been possible without the support of my collaborators: Paul Torrey, Sara Ellison, David Patton, Asa Bluck, Gunjan Bansal and Lars Hernquist. Lastly, I wish to thank Volker Springel for kindly giving me permission to use his \textsc{gadget-3} code, Alexia Lewis for pointing out the M31-M32 encounter and references to it, and the organisers of {\it Galaxies in 3D across the universe} for putting together such an exciting, fruitful, and memorable meeting!

\end{document}